\documentclass[aps,prc,twocolumn,showpacs,groupedaddress,floatfix]{revtex4}

\usepackage{graphicx}
\usepackage{bm}

\begin{document}

\title{Time-Dependent Response Calculations of Nuclear Resonances}

\author{A. S. Umar and V.E. Oberacker}
\affiliation{Department of Physics, Vanderbilt University, Nashville, TN 37235}

\date{\today}

\begin{abstract}
A new alternate method for evaluating linear response theory is
formally developed, and results are presented. This method involves the
time-evolution of the system using TDHF and is constructed
directly on top of a static Hartree-Fock calculation.  By Fourier
transforming the time-dependent result the response function and
the total probability amplitude are extracted.  This method allows
for a coherent description of static properties of nuclei, such as
binding energies and deformations, while also providing a method
for calculating collective modes and reaction rates.  A full
3-D Cartesian Basis-Spline collocation representation is used
with several Skyrme interactions.  Sample results are presented for the
giant multipole resonances of $^{16}$O, $^{40}$Ca, and $^{32}$S
and compared to other calculations.
\end{abstract}
\maketitle

\section{Introduction}

Typically, linear response theory equations are derived by adding
a specific time-dependent perturbing function to the Hamiltonian, which
is usually harmonic in time, resulting in a set of RPA-like
equations. These equations are then solved, for a given energy,
using several methods (see
for example Refs.~\cite{bertsch,krewald,liu,SgII}) to give the
response of the system to a specific collective excitation mode.

In this paper, a new alternate method is presented to calculate
response theory.  A specific time-dependent perturbing
external piece is added to the static Hamiltonian to give a
time-dependent total Hamiltonian, $H_{tot}(t)$.
A static Hartree-Fock solution is then time-evolved using
this $H_{tot}(t)$ in a time-dependent
Hartree-Fock (TDHF) calculation.  The time-dependent result is
then Fourier transformed to give the response of the system for
all energies.  In this scheme one recovers both the response
spectrum as well as the total transition probability amplitude
corresponding to a given specific collective mode. Similar analyses
of the long-time evolution of TDHF equations to study collective 
vibrations have been utilized in the past \cite{BF79,umar1,umar2,SCF01,SC03},
as well as extensions to study the damping of giant resonances \cite{BCT92,TU02}.
The main advantage of this approach is that the dynamical
response calculation is constructed directly on top of a static
Hartree-Fock calculation and hence the static and dynamical
calculations are calculated using the same Hamiltonian
description.  Therefore there is a complete consistency between
the static ground state of the system and the response
calculations.  One can then provide a coherent description of
static properties of nuclei and of dynamical properties.  This is
important for example in $\beta-$decay calculations of exotic
nuclei, where reliable predictions are very sensitive to the
deformation properties of the nucleus \cite{sorlin}.  Hence in
this formalism consistent predictions of both the deformation
and reaction rate properties are possible.

The static and dynamical Hartree-Fock calculations are performed
using a 3-D Cartesian Basis-Spline collocation expansion \cite{umar3,umar4}.
Properties of Basis-Splines include the practicality associated
with coordinate-space lattice grids, while also providing accurate
representations of the gradient operator and a good description
of the continuum.

In Section II,  the time-dependent evaluation of the response theory is
discussed and shown to give the total transition probability.
Numerical details and sample results are presented in Section III.

\section{Time-Dependent Response Theory}

The response equations can be derived from a specific
time-dependent perturbation functional of the
TDHF equations \cite{fetter}.
To begin the proof a solution to the static Schr{\"o}dinger
equation is written as follows:
\begin{equation}
{\widehat H}|\psi_s(0)\rangle = E |\psi_s(0)\rangle . \label{eq:3.1}
\end{equation}
A time-dependent perturbing function is added to the
static Hamiltonian:
\begin{equation}
{\widehat H}_{tot} = {\widehat H} + {\widehat H}_{ex}(t) .
\label{eq:3.2}
\end{equation}
The external piece is defined as:
\begin{eqnarray}
{\widehat H}_{ex}(t) & = & {\widehat F} f(t) \nonumber \\
  & = & \left[ \int d^3x\; {\widehat n}(\bm{x},t) F(\bm{x})\right] f(t) , \label{eq:3.3}
\end{eqnarray}
where ${\widehat n}(\bm{x},t)$ is the number density operator and
$F(\bm{x})$ corresponds to a one-body operator to excite a
particular collective mode.
The functions, $F(\bm{x})$ and $f(t)$ will be chosen later.

At some time $t=t_0$ the external piece of the Hamiltonian is
turned on where $|{\bar \psi}_s(t)\rangle$ is the solution
to the time-dependent Schr\"{o}dinger equation:
\begin{equation}
\imath\hbar\frac{\partial}{\partial t} |{\bar \psi}_s(t)\rangle
 = \left[ {\widehat H} + {\widehat H}_{ex}(t)\right]
     |{\bar \psi}_s(t)\rangle .  \label{eq:3.4}
\end{equation}
Here the subscript ``$s$'' refers to the Schr\"{o}dinger picture.
A solution of the following form is then constructed:
\begin{equation}
|{\bar \psi}_s(t)\rangle = e^{-\imath{\widehat H}t/\hbar}
      {\widehat A}(t) |\psi_s(0)\rangle  , \label{eq:3.5}
\end{equation}
where for $t\leq t_0$, ${\widehat A}(t)=1$.  Using
Eq.~(\ref{eq:3.4}), the function ${\widehat A}(t)$ can be shown
to be a solution to:
\begin{equation}
\imath\hbar\frac{\partial}{\partial t} {\widehat A}(t)
 = {\widehat H}_{ex}^I(t) {\widehat A}(t) , \label{eq:3.6}
\end{equation}
where the superscript ``$I$'' refers to the interaction picture,
which reduces to the Heisenberg picture when
${\widehat H}_{ex}(t)=0$.  The solution to Eq.~(\ref{eq:3.6}) can
be written iteratively as:
\begin{widetext}
\begin{equation}
{\widehat A}(t) = 1 - \frac{\imath}{\hbar}\int_{t_0}^{t} dt'
      {\widehat H}_{ex}^I(t') + ... , \label{eq:3.7}
\end{equation}
where the state vector is then given by:
\begin{equation}
|{\bar \psi}_s(t)\rangle = e^{-\imath{\widehat H}t/\hbar}
 |\psi_s(0)\rangle - \frac{\imath}{\hbar} e^{-\imath{\widehat H}t/\hbar}
 \int_{t_0}^{t} dt'{\widehat H}_{ex}^I(t')|\psi_s(0)\rangle
  + ... ~.\label{eq:3.8}
\end{equation}
The expectation value of any operator, ${\widehat O}(t)$, is
equal to
\begin{eqnarray}
\langle{\bar \psi}_s(t)|{\widehat O}_S(t)|{\bar \psi}_s(t)\rangle
 & = & \langle{\psi}_s(0)|{\widehat O}_I(t)
|{\psi}_s(0)\rangle +  \\
 & ~ & + \langle{\psi}_s(0)|\frac{\imath}{\hbar}
 \int_{t_0}^{t} dt'\left[{\widehat H}_{ex}^I(t'),
{\widehat O}_I(t)\right] |{\psi}_s(0)\rangle + ...~.
      \label{eq:3.9} \nonumber
\end{eqnarray}
The linear approximation is made such that terms beyond first
order in ${\widehat H}_{ex}^I$ are neglected.
The first term in the expansion is trivially the unperturbed expectation
value of the operator in the Schr\"odinger picture.
If we choose the operator ${\widehat O}(t)$ to be the number
density operator, then using Eq.~(\ref{eq:3.3}), the fluctuation
in the density can be defined as:
\begin{eqnarray}
\delta\langle {\widehat n}(\bm{x},t)\rangle & = &
\langle{\bar \psi}_s(t)|{\widehat n}_S(t)|{\bar \psi}_s(t)\rangle
- \langle{\psi}_s(0)|{\widehat n}_S(0)|{\psi}_s(0)\rangle
   \nonumber \\
 & = & \langle{\psi}_s(0)|\frac{\imath}{\hbar}
\int_{t_0}^{t} dt' \int d^3x' F(\bm{x}') f(t')
\left[ {\widehat n}_I(\bm{x}',t'),{\widehat n}_I(\bm{x},t)\right]
 |{\psi}_s(0)\rangle \; . \label{eq:3.10}
\end{eqnarray}
The retarded density correlation function is defined as:
\begin{equation}
\imath D^R(\bm{x},t;\bm{x}',t') = \theta(t-t') \frac{\langle\psi_0 |
 \left[ {\widetilde n}_H(\bm{x}), {\widetilde n}_H(\bm{x}') \right]
  | \psi_0\rangle}{\langle\psi_0 |\psi_0\rangle} , \label{eq:3.11}
\end{equation}
where ${\widetilde n}_H={\widehat n}_H-\langle{\widehat n}_H\rangle$
is the deviation of the number operator in the Heisenberg
picture.  The density fluctuation can be written as:
\begin{equation}
\delta\langle {\widehat n}(\bm{x},t)\rangle = \frac{1}{\hbar}
         \int_{-\infty}^{\infty}
dt' \int d^3x' D^R(\bm{x},t;\bm{x}',t') F(\bm{x}') f(t') . \label{eq:3.12}
\end{equation}
Using the Fourier representation of $\theta(t-t')$, the Fourier
transform of the density correlation function is:
\begin{eqnarray}
\imath D^R(\bm{x},\bm{x}';\omega) & = & \int_{-\infty}^{\infty} d(t-t')
 e^{\imath\omega(t-t')} \imath D^R(\bm{x},t;\bm{x}',t') \nonumber \\
 & = & \sum_n \left\{
\frac{\langle\psi_0| {\widetilde n}_S(\bm{x})|\psi_n\rangle
\langle\psi_n| {\widetilde n}_S(\bm{x}')|\psi_0\rangle }
{\omega - \frac{E_n-E_0}{\hbar} + \imath\eta} -
\frac{\langle\psi_0| {\widetilde n}_S(\bm{x}')|\psi_n\rangle
\langle\psi_n| {\widetilde n}_S(\bm{x})|\psi_0\rangle }
{\omega + \frac{E_n-E_0}{\hbar} + \imath\eta} \right\} ,
  \label{eq:3.13}
\end{eqnarray}
where $|\psi_n\rangle$ represents the full spectrum of the excited
many-body states of $\widehat H$.  The Fourier transform of the density
fluctuation then becomes:
\begin{eqnarray}
\delta\langle {\widehat n}(\bm{x},\omega)\rangle & = &
     \int_{-\infty}^{\infty} dt
  e^{\imath\omega t} \delta\langle n(\bm{x},t)\rangle \\ \nonumber
  & = & \frac{1}{\hbar} \int d^3x' D^R(\bm{x},\bm{x}';\omega)
F(\bm{x}') f(\omega) , \label{eq:3.14}
\end{eqnarray}
where
\begin{equation}
f(\omega)  =  \int_{-\infty}^{\infty} dt'
       e^{\imath\omega t'} f(t') \; .\nonumber
\end{equation}
The linear response structure function, $S(\omega)$ is derived to be:
\begin{eqnarray}
f(\omega) S(\omega) & = & \int d^3x
       \delta\langle F^\dagger(\bm{x}) n(\bm{x},\omega)\rangle \nonumber \\
  & = & \frac{1}{\hbar} \int d^3x \int d^3x' F^\dagger(\bm{x})
      D^R(\bm{x},\bm{x}';\omega) F(\bm{x}') f(\omega)  .
       \label{eq:3.15}
\end{eqnarray}
Combining Eqs.~(\ref{eq:3.13}) and (\ref{eq:3.15}) the imaginary
part of the structure function then gives the total transition
probability associated with $F(\bm{x})$
\begin{equation}
{\it Im} \left[S(\omega)\right] = - \frac{\pi}{\hbar}
    \sum_n \left| \int d^3x'
 \langle\psi_n|{\widetilde n}_x(\bm{x}')|\psi_0\rangle
  F(\bm{x}')\right|^2 \delta\left(\omega -
   \frac{E_n-E_0}{\hbar}\right) , ~~~~~E_n\geq E_0 .\label{eq:3.16}
\end{equation}
Note that this quantity is negative definite; this feature can
be used as a measure of the convergence of the solution.
\end{widetext}
At this point instead of using the standard route of letting
$f(t)\rightarrow 0$ to recover the linear response equations, we
choose an alternate technique to calculate the response.  In this
case we evolve the system in time and then Fourier transform the
result, where $H_{ex}(t)$ is a perturbing function.  We
choose $f(t)$  to be a Gaussian of the following
form,
\begin{eqnarray}
f(t) & = & \varepsilon e^{-\frac{\alpha}{2} (t-t_0)^2},
            ~~~~~t \geq t_0  \nonumber \\
f(\omega) & = & \varepsilon \sqrt{\frac{2\pi}{\alpha}}
           e^{-\frac{\omega^2}{2\alpha}}, \label{eq:3.17}
\end{eqnarray}
where $\varepsilon$ is some small number $(\sim 10^{-6})$, chosen
such that we are in the linear regime.  The parameter, $\alpha$,
is set to be $1.0~{\rm c^2/fm^2}$, which allows for a reasonable
perturbation of collective energies up to $\approx 150$~MeV.

In practice, our numerical calculations proceed as follows:
First, we generate highly accurate static HF wave functions
on the 3-D lattice. Then the external time-dependent perturbation,
Eq.~(\ref{eq:3.3}), is set up by choosing a particular form for
$F(\bm{x})$, and using Eq.~(\ref{eq:3.17}) for $f(t)$.
Next, we solve the TDHF equations utilizing the time-evolution
operator
\begin{equation}
U(t,t_0) = T \left[ e^{-\frac{\imath}{\hbar} \int_{t_0}^{t}
  dt' {\widehat H}_{tot}(t')}\right] , \label{eq:3.18}
\end{equation}
where $T\left[ ~...~\right]$ denotes time-ordering.  Using
infinitesimal time increments, the time-evolution operator is
approximated by
\begin{eqnarray}
U(t_{n+1},t_n) & = & e^{-\frac{\imath}{\hbar} \int_{t_n}^{t_{n+1}}
                   dt' {\widehat H}_{tot}(t')} \nonumber \\
  & \approx & e^{-\frac{\imath}{\hbar} \Delta t
          {\widehat H}_{tot}(t_n+\frac{\Delta t}{2})} \label{eq:3.19} \\
  & \approx & 1 + \sum_{k=1}^{N}\left[ \frac{\left(-\frac{\imath}{\hbar}
            \Delta t {\widehat H}_{tot} \right)^k}{k!}\right] , \nonumber
\end{eqnarray}
where the quantity ${\widehat H}_{tot}^k$ is evaluated by repeated
operations of ${\widehat H}_{tot}$ upon the wave functions.
\begin{figure}[!htb]
\includegraphics*[scale=0.40]{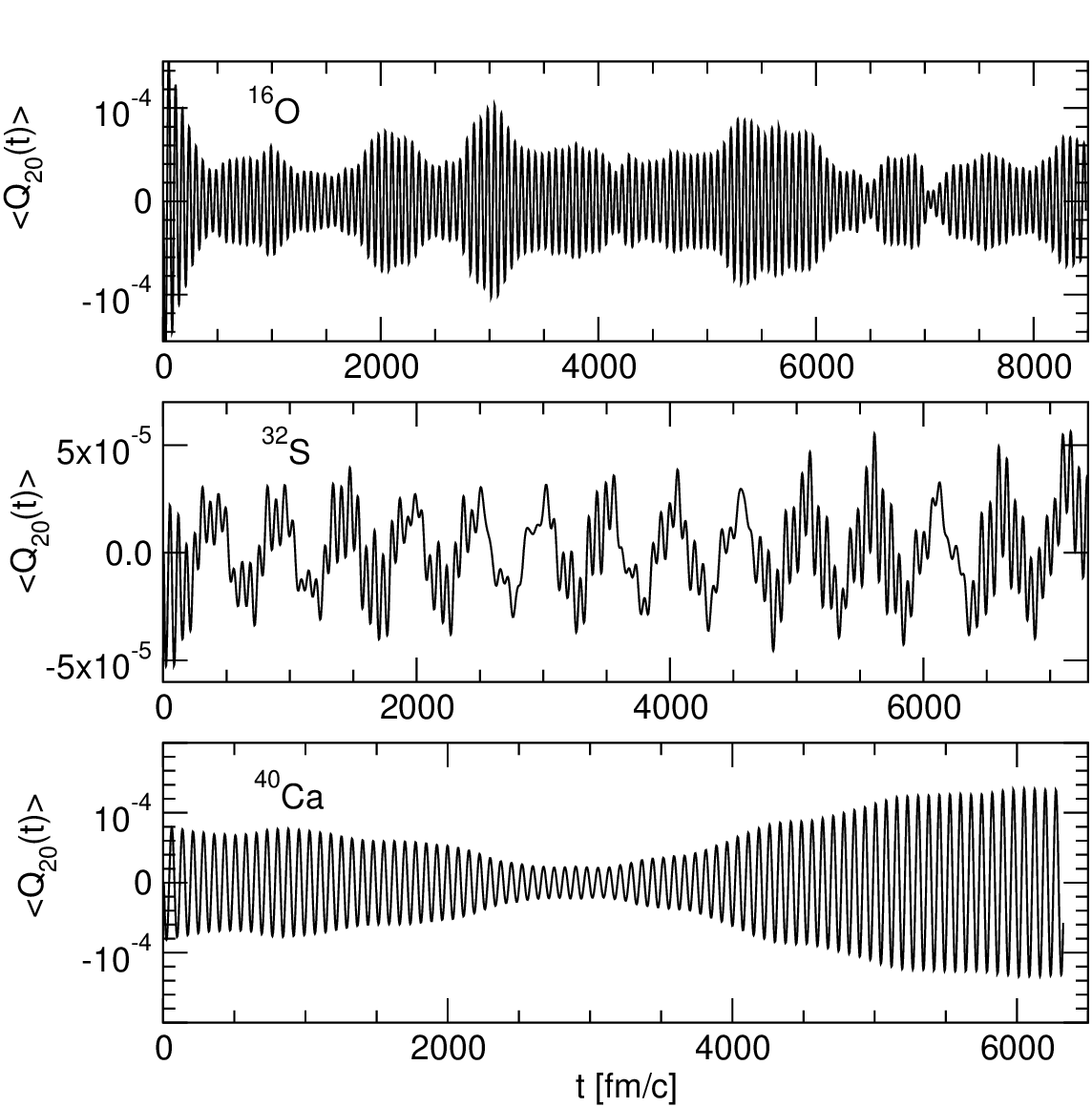}
\caption{\label{fig1}
        The fluctuations in the isoscalar axial quadrupole moment
        as a function of time are shown for $^{16}$O,  $^{32}$S,
	and $^{40}$Ca using the SkM$^*$ interaction. The size
        of the time step is $0.4$~fm/c.  A grid of $24^3$ with a Cartesian box
        dimensioned $(-12, +12~{\rm fm})^3$ is used.}
\end{figure}
From the numerical solution of the TDHF equations, we obtain
the density fluctuation as a function of time, Eq.~(\ref{eq:3.12}).
After Fourier transforming this quantity and folding with the function
$F(\bm{x})$, we obtain $f(\omega) S(\omega)$ in Eq.~(\ref{eq:3.15}), from
which the linear response structure function of the system is
extracted.

\section{Numerical Details and Results}
The static and time-dependent Hartree-Fock calculations are
performed using a collocation spline basis in a three-dimensional
lattice configuration.  Basis-Splines allow for the use of a
lattice grid representation of the nucleus, which is much easier
to use than alternate basis techniques, such as multi-dimensional
harmonic oscillators.  Also, for studies of exotic nuclei, because of
weak binding, the density distributions tend to extend to large
distances and hence one finds a large sensitivity to a harmonic
oscillator basis due to the unphysical description of continuum states,
while for a lattice grid representation one
needs to simply increase the size of the box. Traditional grid
representations typically use finite difference techniques to
represent the gradient operator. Collocation Basis-Splines allow
for the gradient operator to be represented by its action upon a
basis function in a matrix form.  Thus the collocation method gives a
much more accurate representation of the gradient, while
maintaining the convenience of a lattice grid and hence
provides a much more accurate calculation in the end \cite{umar4}.
\begin{figure}[!htb]
\includegraphics*[scale=0.40]{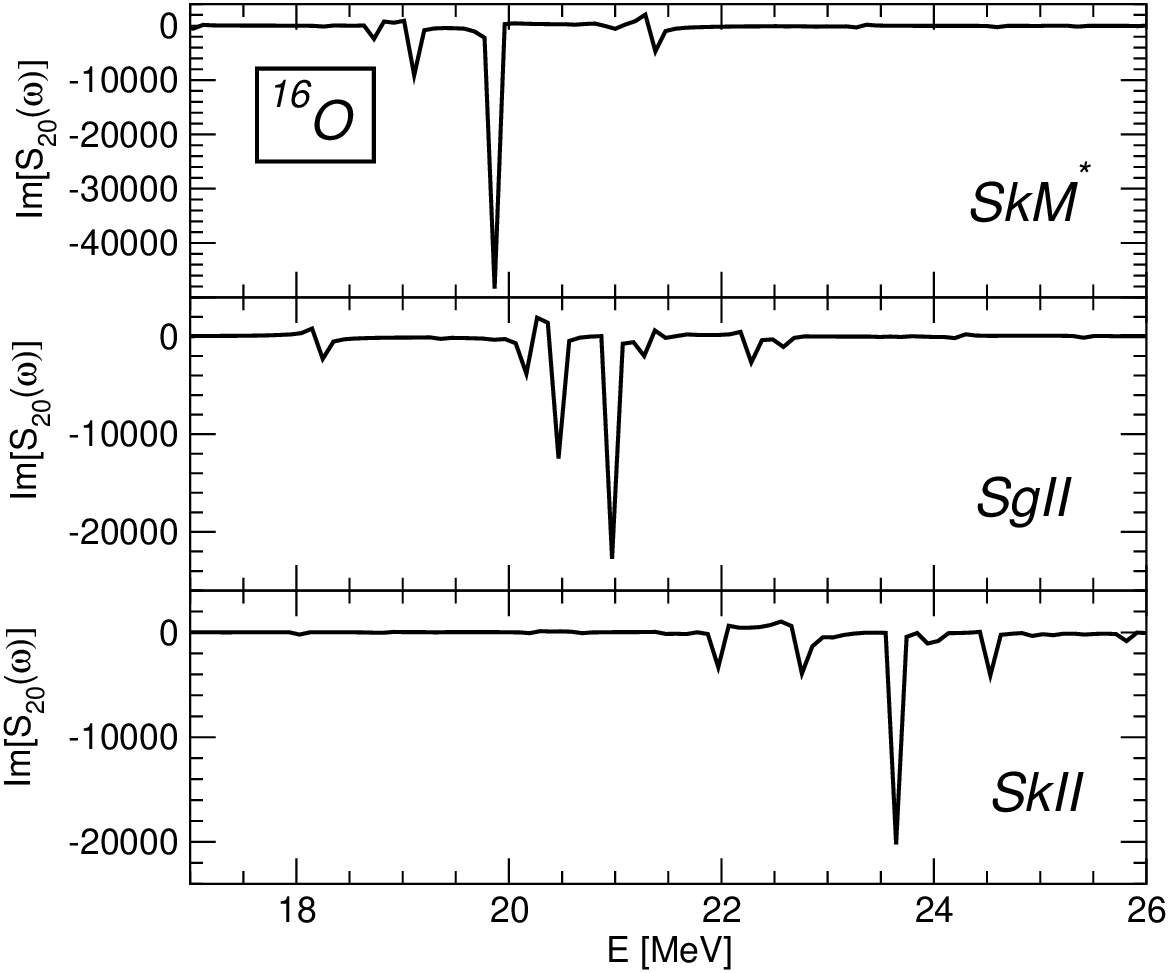}
\caption{\label{fig2}The imaginary part of the response function corresponding
        to the isoscalar quadrupole moment is shown for $^{16}$O using
        three different Skyrme force parametrizations,
        where the SkM$^*$, SgII and SkII versions are shown in the
        upper, middle and lower panels, respectively. The
        experimentally measured giant quadrupole resonance is around 20.7 MeV.}
\end{figure}

We have performed the usual tests for assuring convergence
with respect to the numerical box size and the number of
collocation points.
Typically a 7th order B-spline is used in a
$(12~{\rm fm})^3$ box with $20^3-24^3$ grid points.  
The calculation
can be performed with or without assuming time-reversal symmetry.
The collective linear response may involve particle-hole
interactions which are spin-dependent and not time-reversal
symmetric.  Therefore, for the correct collective content to be
included one should not
impose time-reversal symmetry in the linear response calculations
\cite{krewald}.  In the results presented in this paper
time-reversal symmetry is not imposed.  In a comparison it is
found that imposing time-reversal symmetry causes small shifts in
the position of the collective modes on the order of
$\approx 0.3$~MeV.

Calculations of isovector dipole; isovector and isoscalar
octupole; and isoscalar
quadrupole collective modes are performed for $^{16}$O,  $^{32}$S,
and $^{40}$Ca using several parameterizations of the Skyrme interaction.  Here the
parameterizations known as SkII \cite{SkII}, SkM$^*$ \cite{skm}
and SgII \cite{SgII} are used for comparisons.  The exponent of
the density in the density-dependent term in SkII is $\alpha=1$,
while for SkM$^*$ and SgII this exponent is $\alpha=\frac{1}{6}$.
This causes the SkII force to produce a rather large nuclear matter
incompressibility, while the SkM$^*$ and SgII forces produce more
realistic compression properties.  Also the SkM$^*$ and SgII forces
allow for more stable static Hartree-Fock and TDHF computations.
The static Hartree-Fock calculations converge more easily and
rapidly for SkM$^*$ and SgII than for SkI, SkII or SkIII.
A time step of $\Delta t = 0.4$~fm/c is used for the calculation.
It is found that one can perform the time evolution for up to
32768 time steps without appreciable dissipation.
The results shown here use 16384 time steps for
a maximum time of $\approx 6550$~fm/c.
\begin{figure}[!htb]
\includegraphics*[scale=0.40]{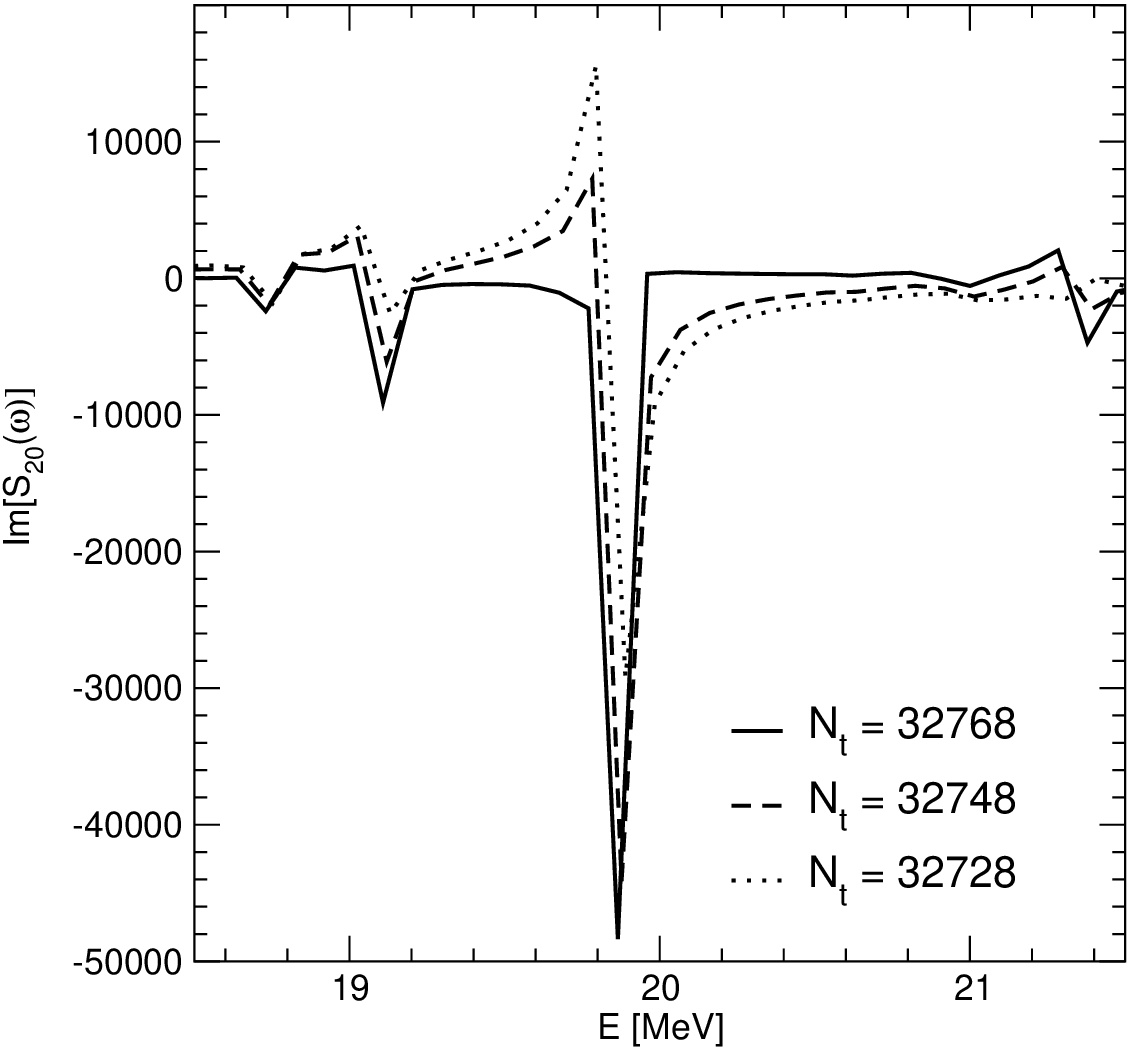}
\caption{\label{fig3} FFT results for $^{16}$O
using three different
numbers of time steps. The case when the number of time steps
equals an exact power of two (32768) is clearly the better
converged result, showing almost no positive values.}
\end{figure}

Reasonable results are obtained if the parameter $\varepsilon$ in
Eq.~(\ref{eq:3.17}) is chosen to fall in the range
$2.0\times 10^{-4}\leq\varepsilon\leq2\times 10^{-7}$.  By varying
the value of $\varepsilon$, the amplitude of the time-dependent
density fluctuation then scales proportionally to $\varepsilon$,
thus indicating that we are well within the linear regime of the
theory.

The linear response calculations require well converged initial
static HF solutions.  To test for the convergence of the static HF
calculation the energy fluctuation, which is the variance of
$\widehat H$, is minimized.  The energy fluctuation is defined as
\begin{equation}
\sqrt{\left|\langle{\widehat H}^2\rangle-\langle
{\widehat H}\rangle^2\right|},
\end{equation}
which measures how close the wave
functions are to being eigenstates of $\widehat H$.  This measure
of convergence is very sensitive to the eigensolutions and is
independent of the iteration step size.  For $^{16}$O it was found
that static HF solutions with energy fluctuation less than
about $1.0\times 10^{-5}$ provided adequate starting points for
the dynamic calculation, although the smaller the energy
fluctuation the better.

The dynamic calculations involve using Eq.~(\ref{eq:3.19}) to
evolve the system.  Since $U(t,t')$ is an unitary operator, the
orthonormality of the system is preserved, therefore it is not
necessary to re-orthogonalize the solutions after every time-step.
The stability of the calculation is checked by testing the
preservation of the norm of each wave function.  The number of
terms in the expansion of the exponent in Eq.~(\ref{eq:3.19}) is
determined by requiring the norm to be preserved to a certain
accuracy (typically to $\leq 1.0\times 10^{-8}-5.0\times 10^{-10}$).
\begin{figure}[!htb]
\includegraphics*[scale=0.40]{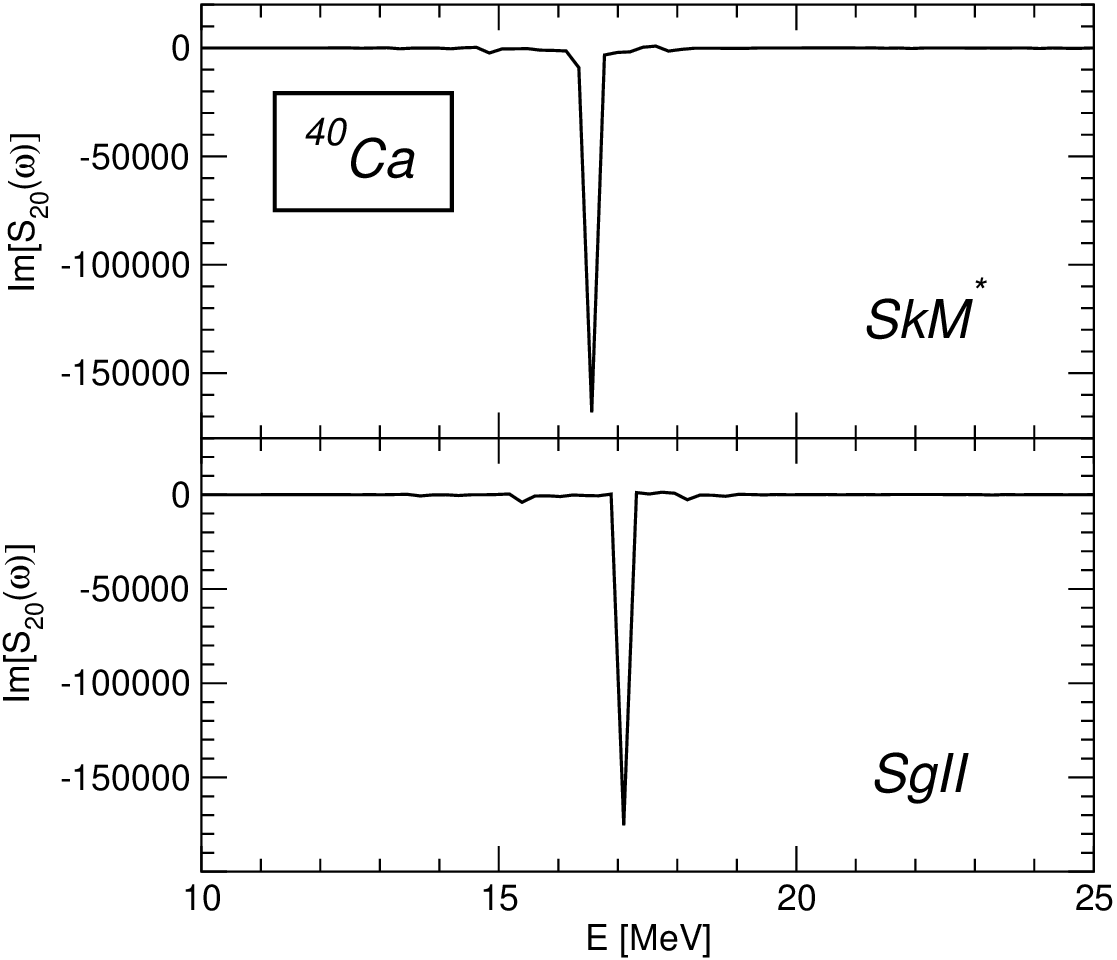}
\caption{\label{fig4}The imaginary part of the response function corresponding
        to the isoscalar quadrupole moment is shown for $^{40}$Ca using
        two different Skyrme force parametrizations,
        SkM$^*$ and SgII.}
\end{figure}

The time-dependent perturbing part of the Hamiltonian is evaluated
when the exponential term in Eq.~(\ref{eq:3.17}) is greater than some
small number, $\varepsilon_{cut}$.  Since it is not
difficult to evaluate the action of
the external part of the Hamiltonian on the wave function,
$\varepsilon_{cut}$ is chosen to be very small, ($1.0\times 10^{-10}$).
To allow the Fourier transform of $f(t)$ to be evaluated easily,
it is necessary to integrate $t$ from $-\infty$ to $\infty$ and
hence we would like the entire Gaussian of the perturbing
function, $f(t)$, to be included into the time evolution to the
desired accuracy.
The parameter $t_0$ is therefore chosen such that the complete
nonzero contribution of the time-dependent perturbation is
included
\begin{equation}
t_0 = -\left( 2\Delta t  + \sqrt{ \left| \frac{2\log{\varepsilon_{cut}}}{\alpha}\right|}\right)\; .
\end{equation}

\subsection{Quadrupole Excitation Modes}

For the study of the isoscalar quadrupole moment, the perturbing
function $F(\bm{x})$, introduced in Eq.~(\ref{eq:3.3}), is
chosen to be the mass quadrupole moment, $Q_{20}=2z^2-(x^2+y^2)$.
It turns out that other even multipole modes are also excited
at the same time ({\it i.e.} $Q_{40}$, $Q_{60}$, ...).
One can therefore study the effect of the coupling between the
different excitation modes.  The same holds true for the odd
multipoles. This is due to the nonlinear response effects
present in the TDHF time evolution \cite{SC03}.

The quadrupole collective resonances are calculated
for $^{16}$O using three different Skyrme force parameterizations
for comparisons.  Two different size grids ($22^3, 24^3$)
were used inside a ($-12,+12$~fm)$^3$ Cartesian
box along with periodic boundary conditions.
\begin{figure}[!htb]
\includegraphics*[scale=0.40]{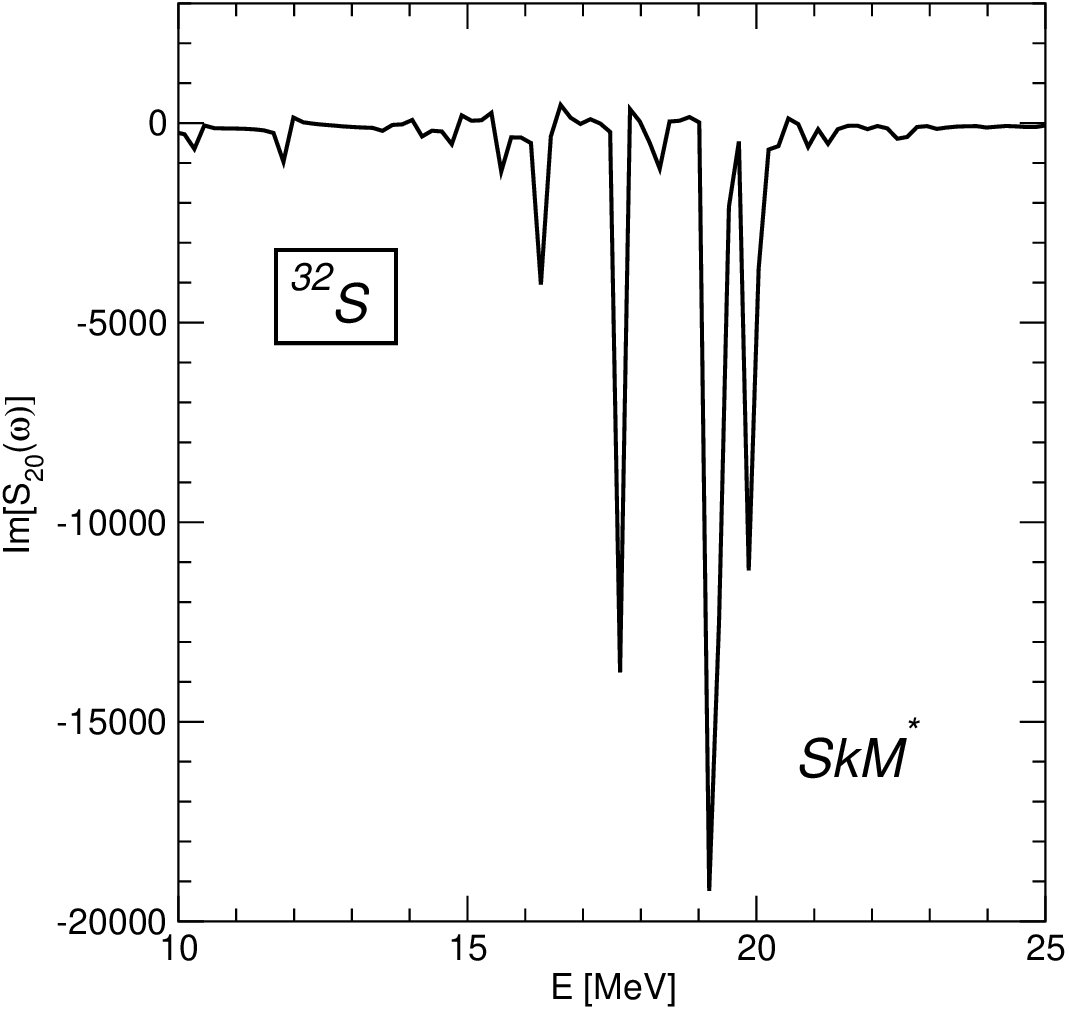}
\caption{\label{fig5}The imaginary part of the response function corresponding
        to the isoscalar quadrupole moment is shown for $^{32}$S using
        the SkM$^*$ force.}
\end{figure}

In Fig. \ref{fig1} the time-dependent evolution of the multipole moment
defined as:
\begin{equation}
\langle {\widehat Q}_{20}(t)\rangle =
\int d^3x \delta\langle{\widehat n}(\bm{x},t)\rangle Q_{20}(\bm{x})\; ,
\label{eq:4.1}
\end{equation}
is shown for  $^{16}$O,  $^{32}$S, and $^{40}$Ca using the SkM$^*$ interaction.
This figure illustrates the periodic
character of the calculations with almost no damping. 
In this case the smallest oscillation is about $65$~fm/c.

A fast Fourier transform (FFT) is used to calculate the Fourier
transform of $\langle{\widehat Q}_{20}(t)\rangle$ to give
$\langle{\widehat Q}_{20}(\omega)\rangle=f(\omega)S_{20}(\omega)$.
The time-dependent perturbation function, $f(\omega)$ can then be
easily factored out using Eq. (\ref{eq:3.15}).  One can use the analytic
expression for $f(\omega)$ or use a numerical FFT calculation,
where the difference between the two methods ends up being negligible.

In Fig. \ref{fig2} the quadrupole responses for $^{16}$O are shown for the
three different Skyrme cases.
The imaginary part of
Eq.~(\ref{eq:4.1}) is Fourier transformed and divided by $f(\omega)$.
Recalling Eq.~(\ref{eq:3.16}), this
quantity is derived to be a negative definite quantity.  The SkM$^*$
results for $^{16}$O reflect this property very well in Fig. \ref{fig2}.
We observe a sharp peak at about $20$~MeV and a response which is
almost purely negative.  The side peaks, which are much less
prominent, may not represent physical effects, but may be due to
the construction of the continuum.
In performing FFT transformations we find it to be important to have
the number of points to be a power of two. In Fig. \ref{fig3} we
demonstrate this by plotting the response function for three different
choices for the total number of time steps which differ from
each other by only 20 points each. As one can see the case for 32768 points
results in almost no positive values, indicating good convergence.
The three cases give similar pole
structures, which are near the experimental isoscalar quadrupole
giant resonance.  The experimental peak is centered at an energy
of $20.7$~MeV with a width of about $7.5\pm 1$~MeV
\cite{data}.  The resonance calculated with the SgII Skyrme
parametrization is closest to the experimental result, although
SkM$^*$ is also fairly close.  The SkII parametrization poorly
represents the data, but this force is known to give a bad
representation of the collective nuclear properties.
However, for the SkII parametrization we were able to compare
our results for $^{16}$O to the continuum RPA calculations of
\cite{krewald}. We find that the difference between the two
calculations is less than 2\% for the L=2, 3, and 4 modes. The
difference may be due to the omission of the spin-orbit term in
the continuum RPA calculations. We find the closest agreement
for the hexadecupole mode, where both calculations give a peak
energy of 28.2 MeV.
The widths of
resonances are not accurately reproduced, because the continuum
is included into the calculation in an approximate fashion and higher
order correlations are missing in the TDHF approach.  Since
the calculations are performed in a box, the continuum is
represented in terms of discrete pseudo-continuum states, whose
density is sensitive to the size of the box.  A larger box size
is expected to better represent the continuum with a higher density of
pseudo-continuum states.

The property represented in Eq.~(\ref{eq:3.16}) of being purely
negative is not strictly reflected by the SgII nor the SkII
results.  One can see that this property is approximately
reflected, but it is clear the convergence of the solution for
these two cases is not nearly as good as for the SkM$^*$ case.
\begin{figure}[!htb]
\vspace*{0.2in}
\includegraphics*[scale=0.40]{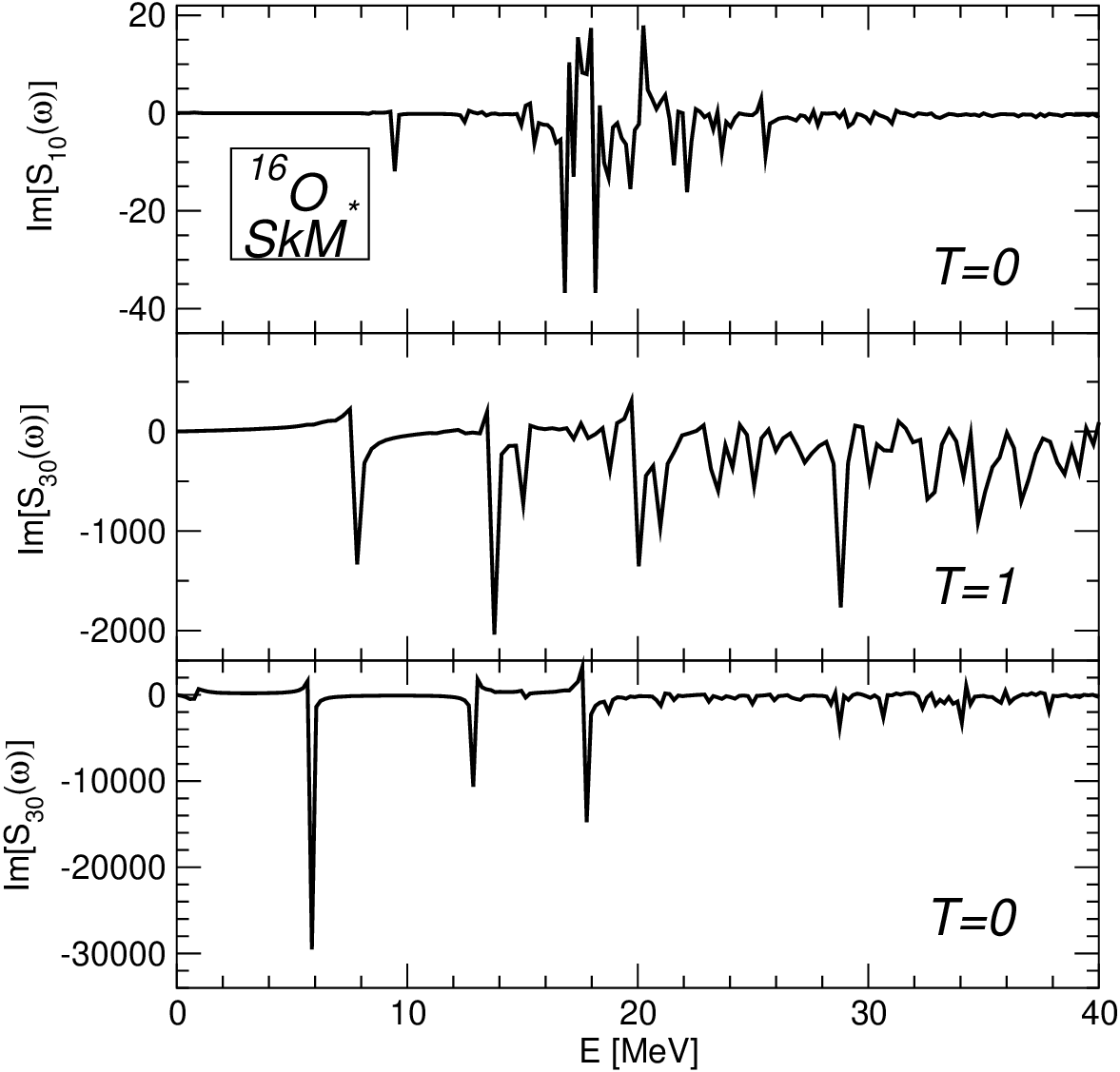}
\caption{\label{fig6}The imaginary part of the response function corresponding
        to the isoscalar and isovector octupole, and isovector dipole are shown for $^{16}$O using
        the SkM$^*$ force.}
\end{figure}

An alternate check of the calculation is the energy-weighted sum
rule (EWSR).  This provides a stringent test of the normalization
and is derivable from the static Hamiltonian.  The SkM$^*$ result
gives $92\%$ of the EWSR, indicating excellent convergence.  For
SgII and SkII the linear response results give $66\%$ and $36\%$
of the EWSR, respectively, indicating a lack of convergence for
these results.

The result for the SgII calculation is particular puzzling, since
it is expected that the SkM$^*$ and SgII interactions should be
similar in behavior.  The violations of the two convergence tests
may be due to coupling to other collective modes, or may be due to
numerical inconsistencies.  We have investigated the SgII
result by varying the size of the box, while keeping the
grid spacing the same.  We have found that a slightly larger box,
(-14,+14) fm, seems to be closer to being purely negative.

In Fig. \ref{fig4} we show the quadrupole strength function for
$^{40}$Ca using two different Skyrme parameterizations. In this
case SkM$^*$ and SgII show a peak at about 17 MeV. We also 
note the strength of the peak. In this case the resonance consumes
all of the EWSR. In Fig. \ref{fig5} we show the same quantity for
$^{32}$S using the Skyrme M$^*$ force. In this case we observe
multiple structures in the giant quadrupole resonance; the energy
splitting arises from the prolate quadrupole deformation of the nuclear
ground state in this system.

The resonant structure of the hexadecupole moment, $Q_{40}$, can
also be calculated at the same time as the quadrupole moment.
This corresponds to the coupling between the quadrupole and the
hexadecupole collective modes.  In this case if one were to
include both $Q_{20}$ and $Q_{40}$ into the external
time-perturbing piece of the total Hamiltonian, then this result
corresponds to the coupling term:

\begin{widetext}
\begin{equation}
f(\omega) S(\omega) = \frac{1}{\hbar} \int d^3x \int d^3x'
F^\dagger_1(\bm{x}) D^R(\bm{x},\bm{x}';\omega)
F_2(\bm{x}') f(\omega).  \label{eq:4.01}
\end{equation}
\end{widetext}
Here $F_1(\bm{x}) \equiv Q_{40}(\bm{x})$ and
$F_2(\bm{x}') \equiv Q_{20}(\bm{x}')$.
For the pure hexadecupole giant resonance both $F_1$ and $F_2$
must be made equal to $Q_{40}$. In general, mixed mode analysis
produces strength functions with less pronounced peaks and can be
used when computational time saving is necessary.
\begin{figure}[!htb]
\includegraphics*[scale=0.40]{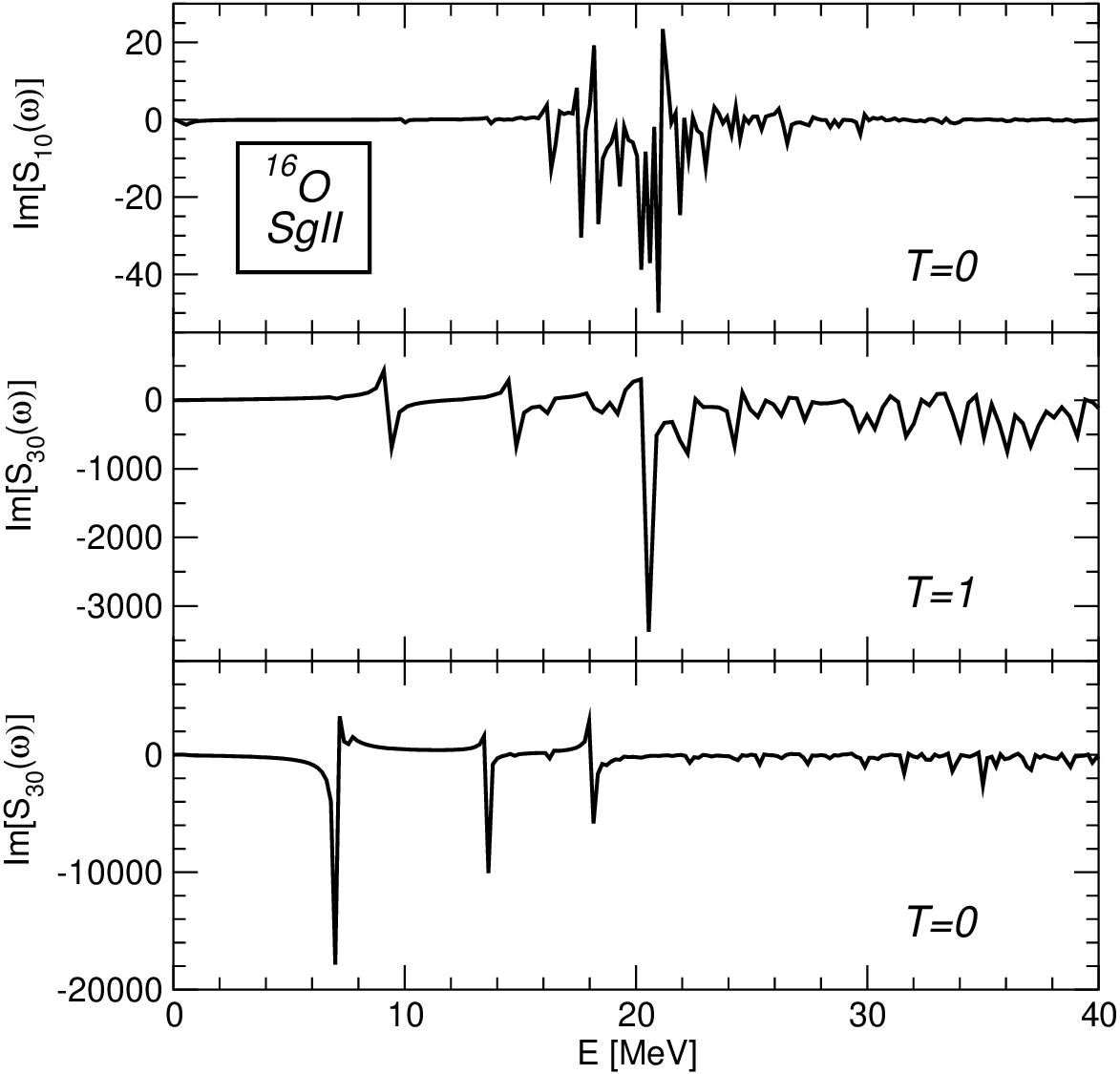}
\caption{\label{fig7}The imaginary part of the response function corresponding
        to the isoscalar and isovector octupole, and isovector dipole are shown for $^{16}$O using
        the SgII force.}
\end{figure}

\subsection{Octupole and Dipole Modes}

The octupole and dipole giant resonances are not symmetric about the
$z = 0$ plane and hence these nodes do not couple with the
symmetric quadrupole and hexadecupole modes.
The isovector octupole moment is defined as:
\begin{equation}
Q_{30} = \frac{1}{2} \left( 1 + \tau_{(3)}\right)
\left[ z^3 - \frac{3}{2} z\left(x^2+y^2\right)\right] .
\end{equation}
In Fig.~6 the isoscalar dipole, isovector octupole, and isoscalar octupole responses are 
shown for  $^{16}$O using the SkM$^*$ interaction. In Fig.~7 the same quantities are
again plotted using the SgII force.
The isoscalar octupole mode shows 
three sharp resonance
structures at about $6-7$~MeV, $13-14$~MeV, and $18$~MeV,
for the SkM$^*$ force, whereas the
SgII resonances are at the slightly higher energies.  These three
peaks are also found in a spherical RPA calculation using the same
effective interactions \cite{PG92}.  The spherical calculation finds
an additional broad peak for SkM$^*$ and SgII centered at an energy
of about $27-28$~MeV.  This peak is weaker than the 3 peaks at
lower energies and is not observed in the three-dimensional linear
response calculation.  In Fig.~6 one can see that the response is
approximately purely negative, but that there are some violations
of this feature.

The isovector octupole response is less clear than the other collective
modes.  It is also possible that resonances seen in the isoscalar
octupole response may also appear in the isovector response due
to couplings. In this case the strength of the peak is expected to
be most prominent in its primary channel. This is most easily observed  in
Fig.~7 for the SgII force. In this  case the
lowest resonances at about $8-9$~MeV and at $14$~MeV are very close to the
lowest peaks in the isoscalar case but with substantially reduced
strength, whereas the most prominent isovector peak is at about
$20$~MeV. The sorting of the isoscalar versus isovector mode is more
complicated for the SkM$^*$ force.

The isovector dipole response resulting from the
linear response calculations is not as prominent as the isoscalar responses.
This is most likely due to the absence of a strong collective dipole resonance
for this nucleus.
For the SkM$^*$ case, there are some peaks centered around
$17-18$~MeV, while for SgII the peaks range about $18-21$~MeV.

\section{Summary and Conclusions}

A method for evaluating the linear response theory using TDHF
is formally developed and implemented.  This method allows one to
construct the dynamic calculation directly on top of the static
Hartree-Fock calculation.  Therefore, by performing a sophisticated
and accurate three dimensional static Hartree-Fock calculation, we
have a correspondingly accurate and consistent dynamic calculation.  A
coherent description of static ground state properties, such as
binding energies and deformations is given along with a
description of the collective modes of nuclei.

A three-dimensional collocation Basis-Spline lattice
representation is used, which
allows for a much more accurate representation of the gradient
operator and hence a correspondingly accurate overall calculation.
Example calculations of two spherical systems ($^{16}$O, $^{40}$Ca)
and of a system with prolate quadrupole deformation ($^{32}$S) are 
presented for the response
functions corresponding to various isoscalar and isovector
multipole moments. The SkM$^*$ case for the axial
isoscalar quadrupole mode gives excellent results, obeying
expectations from the theory and also satisfying the energy
weighted sum rule.

\begin{acknowledgments}
This work is supported by the U.S. Department of Energy under grant No.
DE-FG02-96ER40963 with Vanderbilt University. Some of the numerical calculations
were carried out on IBM-SP supercomputers at the National Energy Research
Scientific Computing Center (NERSC).
\end{acknowledgments}

\end{document}